\documentclass[11pt]{article}
\usepackage{amsfonts,amsthm,amssymb,amsmath}

\newcommand{\di}{\displaystyle}

\newcommand{\al}{\alpha}
\newcommand{\be}{\beta}

\newcommand{\arctg}{\hbox{arctg}}
\newcommand{\sh}{\hbox{sh}}

\newcommand{\ch}{\hbox{ch}}

\newcommand{\ga}{\gamma}

\newcommand{\pa}{\partial}

\newcommand{\R}{{\rm I}\!{\rm R}}

\title{Multitime Rayleigh Solitons}
\vspace{-0.5 cm}
\author{\normalsize Laura Gabriela Matei, Constantin Udriste}
\date{University POLITEHNICA of Bucharest, Faculty of Applied Sciences,
Department of Mathematics-Informatics, Splaiul Independentei 313,
Bucharest 060042, Romania, E-mails: mateiglaura@yahoo.com;
udriste@mathem.pub.ro; anet.udri@yahoo.com}

\begin{document}
\maketitle
\newtheorem{Th}{Theorem}
\newtheorem{Co}{Corollary}
\newtheorem{Prop}{Proposition}

\begin{abstract}
Multitime evolution PDEs for Rayleigh waves are considered, using geometrical ingredients capable to build
an ultra-parabolic-hyperbolic differential operator. Their soliton solutions are found
based on appropriate hypotheses and specific ODEs. These multitime solitons
develop complex behavior of deformation phenomena. The original results include:
the form of multitime Rayleigh PDEs, the construction of multitime Rayleigh solitons 
via some significant amounts of analysis and the stability of  
multitime Rayleigh solitons, which are stable enough to persist indefinitely.
In this context we survey some of the highlights of multitime PDEs theory,
from the more classical single-time case, to the more recent multitime case, 
as well as current developments in using this theory to rigorously prove 
the sense for several evolution variables and PDEs.
\end{abstract}

{\bf Mathematics Subject Classification 2010}: 74J15, 78A50, 74J15, 76Q05.

{\bf Key words}: single-time Rayleigh PDE, multitime Rayleigh PDE, multitime Rayleigh solitons, stability analysis.

\section{Single-time Rayleigh PDEs and our aims}

In the physical and mathematical literature [1]-[4], [7]-[9], we find the {\it Rayleigh wave equation}
$$u_{tt} - u_{xx}=\epsilon (u_t - u_t^3) \eqno(R)$$
related to {\it Rayleigh wave equation of Van der Pol  type}
$$u_{tt} - u_{xx}=\epsilon (1 - u^2)u_t. \eqno(RVP)$$
Each of these has been used to model physical phenomena. Now, the PDE
(R) serves as a model for the large amplitude
vibrations of wind-blown, ice-laden power transmission lines, in time that,
the PDE (RVP) describes plane electromagnetic
waves propagating between two parallel planes in a region
where the conductivity varies quadratically with the electric field.

Just as their counterparts from ordinary differential equations, the PDEs (R) and (RVP)
can be transformed one to another. Their solutions can be obtained by simple
operations performed on the solution of a certain first order, nonlinear wave equation.

An initial-boundary value problem for Rayleigh nonlinear wave equation can be considered to be a simple model
to describe the galloping oscillations of overhead power transmission lines in a wind field. One
end of the transmission line is assumed to be fixed, whereas the other end of the line is assumed
to be attached to a dashpot system.

Rayleigh surface waves are of particular importance in seismology, acoustic, geophysics
and electronics applications.

Some papers (see, [6]) describe as Rayleigh wave a type of seismic surface wave 
that moves with a rolling motion that consists of a combination of particle
motion perpendicular and parallel to the main direction of wave propagation. The amplitude of this motion decreases
with depth. Like primary waves, Rayleigh waves are alternatively compressional and extensional (they cause changes
in the volume of the rocks they pass through).

Section 2 analyzes the geometric objects (fundamental tensor, linear
connection, vector fields, tensor fields) capable of transforming single-time Rayleigh PDEs
into multitime PDEs, showing the existence of an infinity of geometrical structures
such that the multitime Rayleigh PDEs are prolongations of single-time Rayleigh PDEs.
In other words we define an original ultra-parabolic-hyperbolic differential
operators defining the multitime Rayleigh wave equations. Sections 3
underlines the technique which produces multitime Rayleigh
solitons. Sections 4 and 5 praise explicit formulas for the
multitime Rayleigh solitons. Section 6 comments the stability of multitime Rayleigh solitons.

\section{Multitime geometrical prolongations of \\Rayleigh PDEs}

Generally, the passing from systems of PDEs with a single-time variable $t$ to
related PDE systems with $m \geq 2$ evolution variables $t =
(t^\al),\, \al =1,...,m$, is substantially complicated due to
necessity of praising some reasons and some techniques of such
change. The most natural way of changing is to use geometrical
ingredients (derivation, trace etc) that extend the initial system.
The theory and a systematic procedure for the construction of such
new systems is presented here in the context of Rayleigh nonlinear
waves.

This paper provides new results regarding the multitime solitons in two and more temporal dimensions
that can be of interest in physics. We overpass the complexity and, furthermore, the difficulty of performing hand computations
for Rayleigh PDE systems involving many temporal variables by using the symbolic software package in MAPLE.
Our source of inspiration for introducing and studying multitime soliton PDEs is the paper [5]. Also the papers [10]-[17]
contains a lot of ideas in this direction.

Let us introduce and study some {\it multitime geometrical prolongations of the Rayleigh PDEs},
using related connection, fundamental tensor field, vector fields, tensor fields which 
leave on the jet bundle as ingredients in the Differential Geometry of the manifold $\R^m$.

Suppose the {\it multitime} $t=(t^1,...,t^m)\in \R^m$ is a parameter of evolution. 
We endow the manifold (jet bundle of order one)
$J^1(\R\times\R^m, \R\times\R^m)$ with a {\it distinguished symmetric linear connection}
${\Gamma}^\gamma_{\al\beta}={\Gamma}^\gamma_{\al\beta}\left(x, t, u, \di\frac{\pa u}{\pa t}\right)$,
and with a {\it distinguished fundamental symmetric contravariant tensor field}
$h = \left(h^{\al\beta}(x, t, u, \di\frac{\pa u}{\pa t})\right)$ of constant signature $(r,z,s),\,\,r + z + s = m$.
Using a $C^2$ function $ u : \R\times\R^m\to \R$, we build the {\it Hessian operator}
$$
(Hess_{{\Gamma}}u)_{\al\beta} =\frac{\partial ^2 u}{\partial t^\al\partial t^\beta}- {\Gamma} ^\gamma _{\al\beta}\frac{\partial u}{\partial t^\gamma},
\;\;\;\al, \beta, \gamma \in \{1,...,m\}
$$
and its trace, called {\it ultra-parabolic-hyperbolic operator},
$$\square_{{\Gamma},h} u=h^{\al\beta}(Hess_{{\Gamma}}u)_{\al\beta}.$$
We define a {\it multitime PDE} as
$$\square_{{\Gamma},h} u-\frac{\pa ^2u}{\pa x^2}=0,\eqno(1)$$
where $x\in R$ and $t=(t^1,...,t^m)\in R^m$.

- Let $$C^{\ga}(x, t, \eta, \xi), \,\ga =1,...,m$$ be a {\it distinguished vector field} 
and $$B^{\al\be\ga}(x, t, \eta, \xi), \,\al, \be, \ga \in\{1,...,m\}$$
be a {\it distinguished tensor field}. If we adopt the hypothesis
$$h^{\al\beta}(x, t, \eta, \xi){\Gamma}^\gamma_{\al\beta}(x, t, \eta, \xi)\xi_{\gamma}=C^{\gamma}(x, t, \eta, \xi)\xi_{\ga}-B^{\al\be\ga}(x, t, \eta, \xi)\xi_{\al}\xi_{\be}\xi_{\ga},\eqno(2)$$
then we obtain the {\it multitime Rayleigh PDE}
$$h^{\al\beta}\frac{\partial ^2 u}{\partial t^\al\partial t^\beta}-C^{\ga}\frac{\partial u}{\partial t^\ga}+
B^{\al\be\ga}\frac{\partial u}{\partial t^\al}\frac{\partial u}{\partial t^\be}\frac{\partial u}{\partial t^\ga}-\frac{\pa ^2u}{\pa x^2}=0.\eqno(3)$$

- If $C^{\ga}(x, t, \eta, \xi), \,\ga =1,...,m$ and $D^{\ga}(x, t, \eta, \xi), \,  \ga =1,...,m$ 
are  {\it distinguished vector fields}
and the constraint relation is
$$h^{\al\beta}(x, t, \eta, \xi){\Gamma}^\gamma_{\al\beta}(x, t, \eta, \xi)\xi_{\gamma}=C^{\gamma}(x, t, \eta, \xi)\xi_{\ga} - D^{\ga}(x, t, \eta, \xi)\eta ^2\xi_{\ga},\eqno(2^{\prime})$$
then we get a {\it multitime Rayleigh wave equation of Van der Pol  type}
$$h^{\al\beta}\frac{\partial ^2 u}{\partial t^\al\partial t^\beta}-C^{\ga}\frac{\partial u}{\partial t^\ga}+
u^2D^{\ga}\frac{\partial u}{\partial t^\ga}-\frac{\pa ^2u}{\pa x^2}=0.\eqno(3^{\prime})$$

The PDE $(3)$ has two important properties: (i) It is {\it multitime-reversible} 
if and only if $C^\alpha(x,-t) = -C^\alpha(x,t)$,
$B^{\alpha\beta\gamma}(x,-t) = - B^{\alpha\beta\gamma}(x,t)$. In this case, 
the functions $u(x,t)$ and $u(x,-t)$ are solutions of this PDE. 
(ii) It has a stationary solution $u(x)$ if and only if $u(x)$ is solution of the equation
$u^{\prime\prime}(x)=0$. In other words, the stationary solution is $u(x) = ax+b$. 
Geometrically, its graph $(x,t,u(x))$ is a hyperplane in $\R^{1+m+1}$.

The PDE $(3^{\prime})$ has similar properties: (i) It is {\it multitime-reversible} 
if and only if $C^\alpha(x,-t) = -C^\alpha(x,t)$, $D^\alpha(x,-t) = -D^\alpha(x,t)$. 
In this case, the functions $u(x,t)$ and $u(x,-t)$ are solutions of the PDE $(3^{\prime})$.
(ii) It has a stationary solution $u(x) = ax+b$, whose graph
$(x,t,u(x))$ is a hyperplane in $\R^{1+m+1}$.

\begin{Th}
(i) There exists an infinity of geometrical structures ${\Gamma}^\gamma_{\al\beta}$, 
$h^{\al\beta}$, $C^{\ga}$, $B^{\al\be\ga}$ on $\R^m$ such that
a solution of the Rayleigh PDE (R) is also a solution of the multitime Rayleigh PDE (3).

(ii) There exists an infinity of geometrical structures ${\Gamma}^\gamma_{\al\beta}$, 
$h^{\al\beta}$, $C^{\ga}$, $D^{\ga}$ on $\R^m$ such that
a solution of the Rayleigh PDE (RVP) is also a solution of the 
multitime Rayleigh PDE of Van der Pol  type $(3^{\prime})$.
\end{Th}

{\bf Proof} Let $t^1 = t$ and $u = u(x,t^1)$.

(i) Suppose $u = u(x,t^1)$ is a solution of single-time Rayleigh PDE (R). 
The function $v(x,t^1,...,t^m) = u(x,t^1)$ is a solution of the multitime Rayleigh PDE (3)
if the family of geometrical structures 
${\Gamma}^\gamma_{\al\beta}$, $h^{\al\beta}$, $C^{\ga}$, $B^{\al\be\ga}$ is fixed by
$$h^{\al\beta}{\Gamma}^1_{\al\beta}\xi_1=C^1\xi_1 - B^{111}\xi_1\xi_1\xi_1.$$
It is obvious that we have an infinity of geometrical structures that satisfy this algebraic equation.

(ii) Suppose $u = u(x,t^1)$ is a solution of single-time Rayleigh PDE (RVP). 
The function $v(x,t^1,...,t^m) = u(x,t^1)$ is a solution of the multitime Rayleigh PDE $(3^{\prime})$
if the family of geometrical structures ${\Gamma}^\gamma_{\al\beta}$, $h^{\al\beta}$, 
$C^{\ga}$, $D^{\ga}$ is fixed by
$$h^{\al\beta}{\Gamma}^1_{\al\beta}\xi_1=C^1\xi_1 - D^1u^2\xi_1.$$
It is obvious that we have an infinity of geometrical structures that satisfy this algebraic equation.

The foregoing Theorem justifies the term {\it multitime geometrical prolongations of the Rayleigh PDEs}.

Conversely, if we want to obtain a solution of a single-time Rayleigh PDE from a solution of the multitime
Rayleigh PDE, we can use a suitable curve $\tau \to \phi(\tau),\, t^\alpha=\phi^\alpha(\tau),\, \alpha = 1,...,m$,
which imposes some conditions on the coefficients. Particularly, 
we can look for a solution of type $u(x,(\tau,...,\tau))$.

\section{Multitime Rayleigh solitons}

The first aim of this Section is to find some multitime solitons solutions 
for the multitime Rayleigh PDE. In spite of the mathematical beauty,
the distance between theoretical multitime models and real situations where they apply is far from our understanding.

Let $\phi: I\subset \R \to \R$ be a function of class $C^4$.
We seek for solutions of the PDEs (3) and ($3^{\prime}$) in the form of {\it multitime solitons}
$$u(x,t)=\phi (x-\lambda_{\al}t^{\al})=\phi(z),$$
where $\lambda_{\al}$, $\al = 1,...,m$, is a constant vector and $z =x - \lambda_{\al}t^{\al}$.
Then, the partial derivatives of the unknown function $u(x,t)$ are
$$\frac{\pa ^2u}{\pa x^2}=\phi '' (z),\,\,\frac{\pa u}{\pa t^{\al}}=\phi '(z) (-\lambda_{\al}),\,\,
\frac{\pa ^2u}{\pa t^{\al}\pa t^{\be}}=\phi ''(z) \lambda_{\al}\lambda_{\be}.$$
Substituting these derivatives in the PDEs (3) and ($3^{\prime}$), we obtain the second order ODEs,
$$[h^{\al\beta}\lambda_{\al}\lambda_{\be}-1]\,\phi ''(z)-B^{\al\be\ga}\lambda_{\al}\lambda_{\be}\lambda_{\ga}\,\phi '(z)^3+C^{\ga}\lambda_{\ga}\,\phi '(z)=0\eqno(4)$$
and respectively
$$[h^{\al\beta}\lambda_{\al}\lambda_{\be}-1]\,\phi ''(z)-D^{\ga}\lambda_{\ga}\,\phi ^2(z)\phi '(z)+C^{\ga}\lambda_{\ga}\,\phi '(z)=0.\eqno(4^{\prime})$$

Summarizing, we have

\begin{Th} If $\phi (z)$ is a solution of second order ODE
$(4)$ or $(4^{\prime})$ , then $u(x,t)=\phi (x-\lambda_{\al}t^{\al})$ is a multitime soliton solution
of the multitime Rayleigh PDEs (3) or ($3^{\prime}$) respectively.
\end{Th}

In order to find some multitime Rayleigh solitons, we make some particular choices of the
elements that appear in the construction of the two multitime PDEs.

{\bf First choise} We consider the ODE (4) and
we relate the metric tensor $h^{\al \be}(t, x, \eta, \xi)$, the tensor field
$B^{\al\be\ga}(t, x, \eta, \xi)$, the vector field $C^{\ga}(t, x, \eta, \xi)$ 
and the constant vector $\lambda_{\al}$ by the conditions
$$
h^{\al \be}(t, x, \eta, \xi)\lambda_{\al} \lambda_{\be}-1=a(x-\lambda_{\al}t^{\al})= a(z) \not= 0,
$$
$$
B^{\al\be\ga}(t, x, \eta, \xi)\lambda_{\al}\lambda_{\be}\lambda_{\ga}=b(x-\lambda_{\al}t^{\al})= b(z),
$$
$$
C^{\ga}(t, x, \eta, \xi)\lambda_{\ga}= c(x-\lambda_{\al}t^{\al})=c(z).
$$
With these conditions, the ODE (4) becomes
$$a(z)\phi '' - b(z)(\phi ')^3+c(z)\phi '=0.\eqno(5)$$

{\bf Second choise} We take de ODE $(4^{\prime})$
and we relate the metric tensor $h^{\al \be}(t, x, \eta, \xi)$, the vector fields $C^{\ga}(t, x, \eta, \xi)$ and
$D^{\ga}(t, x, \eta, \xi)$ and the constant vector $\lambda_{\al}$ by the conditions
$$
h^{\al \be}(t, x, \eta, \xi)\lambda_{\al} \lambda_{\be}-1=a(x-\lambda_{\al}t^{\al})= a(z) \not= 0,
$$
$$
D^{\ga}(t, x, \eta, \xi)\lambda_{\ga}=b(x-\lambda_{\al}t^{\al})= d(z),
$$
$$
C^{\ga}(t, x, \eta, \xi)\lambda_{\ga}= c(x-\lambda_{\al}t^{\al})=c(z).
$$
With this choice, the ODE $(4^{\prime})$ becomes
$$a(z)\phi '' - d(z)\phi ^2\phi '+c(z)\phi '=0.\eqno(5')$$

We are looking for some solutions of the ODE (5) and $(5^{\prime})$, 
with a view to finding multitime Rayleigh solitons.

\section{Family of multitime Rayleigh solitons}

We start with the multitime Rayleigh solitons based on the ODE (5).

\subsection{Case of coefficients depending on $z$}

Denoting $\phi '\stackrel{\mathrm{not}}{=}\psi$, the second order ODE $(5)$ becomes a first order ODE,
$$a(z)\,\psi ' - b(z)\,\psi ^3+c(z)\,\psi =0,$$
called {\it Bernoulli ODE}. The general form of this ODE is called {\it Abel ODE} of the {\it first kind},
and it arose in the context of the studies of Niels Henrik Abel on
the theory of {\it elliptic functions}, and represents a natural generalization of the Riccati equation.

Since $a(z) \not= 0$, we write the equation in the form
$$\psi '= - \frac{c(z)}{a(z)}\,\psi +\frac{b(z)}{a(z)}\,\psi ^3.$$
By a change of the unknown function, $\xi =\psi ^{-2},$
the Bernoulli ODE becomes a linear ODE,
$$\xi '-2\,\frac{c(z)}{a(z)}\,\xi=-2\,\frac{b(z)}{a(z)},$$
with the solutions
$$\xi(z)=\exp\left({2\di\int\frac{c(z)}{a(z)}dz}\right)\, 
\left[K-2\int\frac{b(z)}{a(z)}\, \exp\left({-2\di\int\frac{c(z)}{a(z)}dz}\right)dz\right], \,\, K\in \R.$$
Since $\psi =\xi^{-\frac{1}{2}}$, we obtain
$$\psi(z)=\exp\left({-\di\int\frac{c(z)}{a(z)}dz}\right)\, \left[K-2\int\frac{b(z)}{a(z)}\, \exp\left({-2\di\int\frac{c(z)}{a(z)}dz}\right)dz\right]^{-\frac{1}{2}}, \,\, K\in \R,$$
that is
$$\phi '(z)=\frac{\exp\left({-\di\int\frac{c(z)}{a(z)}dz}\right)}{\sqrt{K-2\di\int\frac{b(z)}{a(z)}\, \exp\left({-2\di\int\frac{c(z)}{a(z)}dz}\right)dz}}, \,\, K\in \R.$$
Therefore, we have found solutions of the multitime  PDE Rayleigh (3):\\

\begin{Th} If we fix the above coefficients by the conditions
$$
h^{\al \be}(t, x, \eta, \xi)\lambda_{\al} \lambda_{\be}-1= a(x-\lambda_{\al}t^{\al}) \not= 0,
$$
$$
B^{\al\be\ga}(t, x, \eta, \xi)\lambda_{\al}\lambda_{\be}\lambda_{\ga}= b(x-\lambda_{\al}t^{\al}),
C^{\ga}(t, x, \eta, \xi)\lambda_{\ga}= c(x-\lambda_{\al}t^{\al}),
$$
then the function
$$u(x,t)=\phi (x-\lambda_{\al}t^{\al})$$ represents a multitime soliton-solution 
for the multitime PDE Rayleigh (3) for every $\phi$ given by
$$\phi(z)=\int\frac{\exp\left({-\di\int\frac{c(z)}{a(z)}\,dz}\right)}{\sqrt{K-2\di\int\frac{b(z)}{a(z)}\, \exp\left({-2\di\int\frac{c(z)}{a(z)}\,dz}\right)dz}}\,dz, \,\, K\in \R.$$
\end{Th}

\subsection{Case of constant coefficients}

If we fix the elements $h, B, C, \lambda $ by the constants
$$
h^{\al \be}(t, x, \eta, \xi)\lambda_{\al} \lambda_{\be}-1=a \not= 0,
$$
$$
B^{\al\be\ga}(t, x, \eta, \xi)\lambda_{\al}\lambda_{\be}\lambda_{\ga}= b,\,\,
C^{\ga}(t, x, \eta, \xi)\lambda_{\ga}= c,
$$
then the ODE (5) takes the form
$$a\phi '' - b(\phi ')^3+c\phi '=0.$$
Denoting $\phi '\stackrel{\mathrm{not}}{=}\psi$, this second order ODE with constant coefficients
becomes a first order {\it Bernoulli ODE}
$$a\,\psi '-b\,\psi ^3+c\,\psi =0.$$
Having separable variables, this ODE can be written
$$a\, \frac{d\psi}{b\psi^3-c\psi}=dz\,\,\,\Leftrightarrow\,\,\, \frac{a}{b}\int \frac{d\psi}{\psi \left(\psi ^2-\di\frac{c}{b}\right)}=\int dz$$
and we get the next sequence of equivalences
$$\frac{a}{c}\left(-\int \frac{d\psi}{\psi}+\int\frac{\psi d\psi}{\psi ^2-\di\frac{c}{b}}\right)=\int dz
\Leftrightarrow \ln \left(\frac{\sqrt{\left|\psi ^2-\di\frac{c}{b}\right|}}{|\psi|}\right)=\frac{c}{a}z+l, \;\; l\in\R$$
$$\Leftrightarrow \psi ^2-\frac{c}{b}=p\,\psi^2 \exp(\frac{2c}{a}z)
\Leftrightarrow \psi(z)=\pm\sqrt{\frac{c}{b(1-p\, \exp(\frac{2c}{a}z))}}, \;\;p\in\R^*$$
that is
$$\phi(z)=\pm\int \sqrt{\frac{c}{b(1-p\, \exp(\frac{2c}{a}z))}}\; dz, \;\;p\in\R^*,$$ 
with $\frac{c}{b(1-p\,\exp(\frac{2c}{a}z))}\geq 0$.
To calculate the primitive from the right hand side, we amplify the fraction by 
$\exp({\frac{2c}{a}z})$ and then make the change of variables
$\exp(\frac{c}{a}z)=t$. We obtain the integral
$$I = \pm\, \frac{a}{c}\int\frac{1}{t}\sqrt{\frac{c}{b(1-pt^2)}}\; dt$$
and a new change of variables, $\di\frac{1}{t}=s$, gives a new integral,
$$J=\pm\, \frac{a}{c}\int \sqrt{\frac{c}{b(s^2-p)}}\; ds.$$
There are two separate possibilities:

a) If we take $\di\frac{c}{b}>0$, then
$$J=\pm\, \frac{a}{c}\sqrt{\frac{c}{b}}\int \frac{1}{\sqrt{s^2-p}}\; ds.$$
- For $p>0$, the integral is
$$J=\pm\, \frac{a}{c}\sqrt{\frac{c}{b}}\, \ch^{-1}\left|\frac{s}{\sqrt{p}}\right|+r, \;\; r\in\R,$$
that is
$$\phi(z)=\pm\, \frac{a}{c}\sqrt{\frac{c}{b}}\, \ch^{-1}\left|\frac{1}{\sqrt{p}\exp(\frac{c}{a}z)}\right|+r, \;\; r\in\R,\;\; p\in\R^*_+$$
and we can write
$$\phi(z)=\pm\, \frac{a}{c}\sqrt{\frac{c}{b}}\, \ch^{-1}\left(K\exp({-\di\frac{c}{a}z})\right)+r, \;\; r\in\R,\;\; K\in\R^*_+.$$
- For $p<0$, we have
$$J=\pm\, \frac{a}{c}\sqrt{\frac{c}{b}}\int \sqrt{\frac{1}{s^2+p'}}\; ds, \;\; p'= - p,\;\; p'>0.$$
It follows
$$J=\pm\, \frac{a}{c}\sqrt{\frac{c}{b}}\,\, \sh^{-1}\left|\frac{s}{\sqrt{p'}}\right| + r, \;\; r\in\R,$$
that is
$$\phi(z)=\pm\, \frac{a}{c}\sqrt{\frac{c}{b}}\,\, \sh^{-1}\left|\frac{1}{\sqrt{p'}\exp(\frac{c}{a}z)}\right|+r, \;\; r\in\R,\;\; p'\in\R^*_+$$
and then
$$\phi(z)=\pm \frac{a}{c}\sqrt{\frac{c}{b}}\,\, \sh^{-1}\left(K\exp({-\di\frac{c}{a}z})\right)+r, \;\; r\in\R,\;\; K\in\R^*_+.$$

b) If we suppose $\di\frac{c}{b}<0$, then, via $\di\frac{c}{b(1-p\,  \exp(\frac{2c}{a}z))}\geq 0$, it follows $p>0$ and the integral $J$ becomes
$$J=\pm\, \frac{a}{c}\sqrt{\frac{-c}{b}}\int \frac{1}{\sqrt{p-s^2}}\; ds, \;\; p>0$$
$$\Leftrightarrow J=\pm\, \frac{a}{c}\sqrt{\frac{-c}{b}}\, \arcsin\left(\frac{s}{\sqrt{p}}\right)+r, \;\; r\in\R,\;\; p\in\R^*_+.$$
It follows
$$\phi(z)=\pm\, \frac{a}{c}\sqrt{\frac{-c}{b}}\, \arcsin\left(\frac{1}{\exp(\frac{c}{a}z)\sqrt{p}}\right)+r, \;\; r\in\R,\;\; p\in\R^*_+,$$
that is
$$\phi(z)=\pm\, \frac{a}{c}\sqrt{\frac{-c}{b}}\, \arcsin\left(K\exp({-\di\frac{c}{a}z})\right)+r, \;\; r\in\R,\;\; K\in\R^*_+.$$

Therefore, we have found three families of solutions of the multitime Rayleigh PDE (3):\\

\begin{Th} a) If we fix the above coefficients by the conditions
$$h^{\al \be}\lambda_{\al} \lambda_{\be}=\hbox{constant} ,\;\;\;\;\; Q = \frac{C^{\ga}\lambda_{\ga}}{B^{\al\be\ga}\lambda_{\al}\lambda_{\be}\lambda_{\ga}}>0,$$
then we get two families of soliton solutions
$$u(x,t)=\pm\frac{h^{\al \be}\lambda_{\al} \lambda_{\be}-1}{C^{\ga}\lambda_{\ga}}\sqrt{Q}
\, \ch^{-1}\!\left(\!K\exp\left({-\di\frac{C^{\ga}\lambda_{\ga}}{h^{\al \be}\lambda_{\al} \lambda_{\be}-1}(x-\lambda_{\al}t^{\al})}\right)\!\right)+r,$$
$$u(x,t)=\pm\frac{h^{\al \be}\lambda_{\al} \lambda_{\be}-1}{C^{\ga}\lambda_{\ga}}\sqrt{Q}
\, \sh^{-1}\!\left(\!K\exp\left({-\di\frac{C^{\ga}\lambda_{\ga}}{h^{\al \be}\lambda_{\al} \lambda_{\be}-1}(x-\lambda_{\al}t^{\al})}\right)\!\right)+r,$$
where $r\in\R,\;\; K\in\R^*_+$;

b) if we take
$$h^{\al \be}\lambda_{\al} \lambda_{\be}=\hbox{constant},\;\;\;\;\; Q=\frac{C^{\ga}\lambda_{\ga}}{B^{\al\be\ga}\lambda_{\al}\lambda_{\be}\lambda_{\ga}}<0,$$
then we have another family of soliton solutions,
$$u(x,t) = \pm\,\frac{h^{\al \be}\lambda_{\al} \lambda_{\be}-1}{C^{\ga}\lambda_{\ga}}\,\sqrt{-Q}$$
$$\times \arcsin\!\left(\!K\exp\left({-\di\frac{C^{\ga}\lambda_{\ga}}{h^{\al \be}\lambda_{\al} \lambda_{\be}-1}(x-\lambda_{\al}t^{\al})}\right)\!\right)+r,$$
where $r\in\R,\;\; K\in\R^*_+$.
\end{Th}

\subsection{Mac-Laurin series soliton of \\multitime Rayleigh PDE}

In a previous section, we have obtained a second order ODE in $\phi$, namely,
$$a(z)\phi ''-b(z)(\phi ')^3+c(z)\phi '=0,$$
One assume that it has a solution which is 
analytic on an interval around $z=0$ and we search a Mac-Laurin series solution.
Then we express $\phi$ as a power series in the form
$$
\phi(z) = \sum^{\infty}_{n = 0} \,\al_n\,z^n\eqno(6)
$$
and we try to determine what the $\al_n$'s need to be.
The resulting power series need to converge on an interval around origin.

We compute $\phi '(z)$ and $\phi '^3(z)$and $\phi ''(z)$:
$$\phi '(z) = \sum^{\infty}_{n = 1} \,n\al_n\,z^{n-1}=\sum^{\infty}_{n = 0} \,(n+1)\al_{n+1}\,z^n=\sum^{\infty}_{n = 0} \,\be_{n}\,z^n,$$
$$(\phi ') ^2(z) =\sum^{\infty}_{n = 0} \,\left(\sum ^{n}_{k=0}\!\be _{k}\be _{n-k}\right) \,z^n=$$
$$  =\sum^{\infty}_{n = 0} \,\left(\sum ^{n}_{k=0}\! \al_{k+1}\,\al_{n-k+1}(k+1)(n-k+1)\right) \,z^n= \sum^{\infty}_{n = 0} \,\ga _{n} \,z^n,$$
$$(\phi ')^3(z) = \sum^{\infty}_{n = 0} \,\left(\sum ^{n}_{i=0}\! \ga_i\,\be_{n-i}\right) \,z^n=$$
$$=\sum^{\infty}_{n = 0} \,\left(\sum ^{n}_{i=0} \sum ^{i}_{k=0}\al_{k+1}\,\al_{i-k+1}\,\al_{n-i+1}(k+1)(i-k+1)(n-i+1)\right) \,z^n,$$
$$\phi ''(z) = \sum^{\infty}_{n = 1} \,n(n+1)\al_{n+1}\,z^{n-1}=\sum^{\infty}_{n = 0} \,(n+1)(n+2)\al_{n+2}\,z^n.$$
Consider the particular case
$$a(z)=mz+a, \,\, b(z)=pz+b, \,\, c(z)=qz+c, \,\,\, m,p,q,a,b,c\in \R,$$
that is the coefficients $a(z), b(z), c(z)$ of the ODE (5) are affine functions in $z$.
Consequently, the foregoing ODE gives the identity
$$(mz+a)\sum^{\infty}_{n = 0} \,(n+2)(n+1)\al_{n+2}\,z^n+(qz+c)\sum^{\infty}_{n = 0} \,\al_{n+1}(n+1)\,z^n-$$
$$-(pz+b)\sum^{\infty}_{n = 0} \,\left(\sum ^{n}_{i=0} \sum ^{i}_{k=0}\al_{k+1}\,\al_{i-k+1}\,\al_{n-i+1}(k+1)(i-k+1)(n-i+1)\right) \,z^n=0.$$
This identity can be written
$$m\sum^{\infty}_{n = 1} \,n(n+1)\al_{n+1}\,z^n+2a\al_ 2+a\sum^{\infty}_{n = 1} \,(n+2)(n+1)\al_{n+2}\,z^n-$$
$$-p\sum^{\infty}_{n = 1} \,\left(\sum ^{n-1}_{i=0} \sum ^{i}_{k=0}\al_{k+1}\,\al_{i-k+1}\,\al_{n-i}(k+1)(i-k+1)(n-i)\right) \,z^n-b\al _1 ^3-$$
$$-b\sum^{\infty}_{n = 1} \,\left(\sum ^{n}_{i=0} \sum ^{i}_{k=0}\al_{k+1}\,\al_{i-k+1}\,\al_{n-i+1}(k+1)(i-k+1)(n-i+1)\right) \,z^n+$$
$$+q\sum^{\infty}_{n = 1} \, n\al_n\,z^n+c\al _1+c\sum^{\infty}_{n = 1} \,\al_{n+1}(n+1)\,z^n=0,$$
or, equivalent,
$$(2a\al_ 2+c\al _1-b\al _1 ^3)+\sum^{\infty}_{n = 1}[mn(n+1)\al_{n+1}+a(n+2)(n+1)\al_{n+2}-$$
$$-p\sum ^{n-1}_{i=0} \sum ^{i}_{k=0}\al_{k+1}\,\al_{i-k+1}\,\al_{n-i}(k+1)(i-k+1)(n-i)-$$
$$-b\sum ^{n}_{i=0} \sum ^{i}_{k=0}\al_{k+1}\,\al_{i-k+1}\,\al_{n-i+1}(k+1)(i-k+1)(n-i+1)+$$
$$+qn\al_n+c(n+1)\al_{n+1}]z^n=0.$$
By identifying the coefficients of the powers of $z$ with $0$, we find the condition
$$2a\al_ 2+c\al _1-b\al _1 ^3=0$$
and the recurrence
$$mn(n+1)\al_{n+1}+a(n+2)(n+1)\al_{n+2}-$$
$$-p\sum ^{n-1}_{i=0} \sum ^{i}_{k=0}\al_{k+1}\,\al_{i-k+1}\,\al_{n-i}(k+1)(i-k+1)(n-i)-$$
$$-b\sum ^{n}_{i=0} \sum ^{i}_{k=0}\al_{k+1}\,\al_{i-k+1}\,\al_{n-i+1}(k+1)(i-k+1)(n-i+1)+$$
$$+qn\al_n+c(n+1)\al_{n+1}=0, \;\;\; n\geq 1.\eqno(7)$$
By the initial conditions
$\phi (0)=\al_0$,  $\phi '(0)=\al_1$ and  $2a\al_ 2+c\al _1-b\al _1 ^3=0$, 
this recurrence gives us all the coefficients of the power series (6),
but the difficult part is just solving the recurrence for the unknown $\al_n$.

\begin{Th} In the foregoing hypotheses, the multitime series soliton solution of the multitime Rayleigh PDE is
$$u(x, t) = \sum^{\infty}_{n = 0}\al _n (x - \lambda_{\al}t^{\al})^n,$$
with $\al_0, \, \al_1$ fixed, $2a\al_ 2+c\al _1-b\al _1 ^3=0$ and $\al_n, \;\; n\geq 2$ given by the recurrence (7).
\end{Th}

\section{Family of multitime Rayleigh solitons of \\Van der Pol type}

Now we continue with the multitime Rayleigh solitons of Van der Pol type based on the ODE ($7^{\prime}$).

\subsection{Case of coefficients depending on $z$}

As we have seen in Section 3, if we can choose the constant vector $\lambda_{\al}$ by some conditions
 which relate the elements $h, C, D, \lambda$, then we get the equation
$$a(z)\phi ''(z) - d(z)\phi ^2(z)\phi '(z)+c(z)\phi '(z)=0.$$
We can write this equation in the form
$$\phi ^2(z)\phi '(z)=\frac{a(z)}{d(z)}\phi ''(z)+\frac{c(z)}{d(z)}\phi '(z).$$
In order to find some solutions of this equation, we make a particular choice for $\lambda_{\al}$, by a new  condition:
the coefficients $a(z)\stackrel{\mathrm{not}}{=}h^{\al \be}(t, x, \eta, \xi)\lambda_{\al} \lambda_{\be}-1$,
$d(z)\stackrel{\mathrm{not}}{=}D^{\ga}(t, x, \eta, \xi)\lambda_{\ga}$ and
$c(z)\stackrel{\mathrm{not}}{=}C^{\ga}(t, x, \eta, \xi)\lambda_{\ga}$ will be related by the equality
$$\left(\frac{a(z)}{d(z)}\right)'=\frac{c(z)}{d(z)},$$
or equivalent
$$a'(z)d(z)-a(z)d'(z)=d(z)c(z).$$
With such a selection of $\lambda_{\al}$, the equation becomes
$$\phi ^2(z)\phi '(z)=\left(\frac{a(z)}{d(z)}\phi '(z)\right)',$$
that is
$$\frac{\phi ^3(z)}{3}=\frac{a(z)}{d(z)}\phi '(z)+k, \,\,\, k\in \R.$$
This Bernoulli ODE has separable variables. By integration, we get
$$\int\frac{3d\phi}{\phi ^3-3k}=\int \frac{d(z)}{a(z)}\,dz, \,\,\, k\in \R.\eqno(8)$$
For simplicity, we can take $k=0$ and we find a particular family of solutions, defined by the relation
$$\phi ^2(z)=-\frac{2}{3}\int \frac{d(z)}{a(z)} \, dz.$$
If we keep $k$ variable and non-zero, then making the substitution
$3k$ by $k_1^3$,  the equality (8) becomes
$$\int\frac{3d\phi}{\phi ^3-k_1^3}=\int \frac{d(z)}{a(z)}\,dz, \,\,\, k_1\in \R^*.$$
The integral from the left hand is
$$\int\frac{3d\phi}{(\phi -k_1)(\phi ^2+\phi k_1+k_1^2)}=\frac{1}{k_1^2}\int \frac{d\phi}{\phi-k_1}-\frac{1}{k_1^2}\int \frac{\phi +2k_1}{\phi ^2+\phi k_1+k_1 ^2}\, d\phi$$
$$=\frac{1}{k_1^2}\ln{|\phi -k_1|}-\frac{1}{2k_1^2}\int \frac{2\phi +k_1}{\phi ^2+\phi k_1+k_1 ^2}\, d\phi -\frac{3}{2k_1}\int \frac{d\phi}{\left(\phi +\di\frac{k_1}{2}\right)^2+\di\frac{3k_1^2}{4}}$$
$$=\frac{1}{k_1^2}\ln{|\phi -k_1|}-\frac{1}{2k_1^2}\ln(\phi ^2+\phi k_1+k_1 ^2)-\frac{\sqrt{3}}{k_1^2}\arctg \left(\frac{2\phi +k_1}{k_1\sqrt{3}}\right)+c_1, \,\,\, c_1\in \R.$$
Therefore, the general solution of the equation (8) is expressed implicitly by the equality
$$\frac{1}{k_1^2}\ln{\frac{|\phi -k_1|}{\sqrt{\phi ^2+\phi k_1+k_1 ^2}}}-\frac{\sqrt{3}}{k_1^2}\arctg \left(\frac{2\phi +k_1}{k_1\sqrt{3}}\right)=\int \frac{d(z)}{a(z)}\,dz, \,\,\, k_1\in \R^*.$$

Summarizing, we can formulate the next result:

\begin{Th} If we can take the constant vector $\lambda_{\al}$ so as $h^{\al \be}(t, x, \eta, \xi)$, $B^{\ga}(t, x, \eta, \xi)$,
 $C^{\ga}(t, x, \eta, \xi)$ and $\lambda_{\al}, \,\, \al, \be, \ga =1,...,m$, to be related by the conditions
 $$
h^{\al \be}(t, x, \eta, \xi)\lambda_{\al} \lambda_{\be}-1=a(x-\lambda_{\al}t^{\al}) \not= 0,
B^{\ga}(t, x, \eta, \xi)\lambda_{\ga}=b(x-\lambda_{\al}t^{\al}),
$$
$$
C^{\ga}(t, x, \eta, \xi)\lambda_{\ga}= c(x-\lambda_{\al}t^{\al}), a'(z)d(z)-a(z)b'(z)=d(z)c(z).
$$
then the function $u(x,t)=\phi (x-\lambda_{\al}t^{\al})$ represents a multitime soliton-solution
for the multitime PDE Van der Pol ($3^{\prime}$), for every $\phi$ defined implicitly by one of the equalities
$$\phi ^2(z)=-\frac{2}{3}\int \frac{d(z)}{a(z)} \, dz$$
or
$$\frac{1}{k_1^2}\ln{\frac{|\phi(z) -k_1|}{\sqrt{\phi ^2(z)+\phi(z) k_1+k_1 ^2}}}-
\frac{\sqrt{3}}{k_1^2}\arctg \left(\frac{2\phi(z) +k_1}{k_1\sqrt{3}}\right)=\int \frac{d(z)}{a(z)}\,dz, \,\,\, k_1\in \R^*.$$
\end{Th}

\subsection{Case of constant coefficients}

If we fix the elements $h, C, D, \lambda $ by the constants
$$
h^{\al \be}(t, x, \eta, \xi)\lambda_{\al} \lambda_{\be}-1=a \not= 0,\,
D^{\ga}(t, x, \eta, \xi)\lambda_{\ga}= d,\,
C^{\ga}(t, x, \eta, \xi)\lambda_{\ga}=c,\eqno (9)
$$
then the ODE ($7^{\prime}$) takes the form
$$a\,\phi '' - d\,\phi ^2\phi '+c\,\phi '=0.$$
Integrating, we find a first order ODE, with separable variables
$$a\,\phi ' - \frac{d}{3}\,\phi ^3+c\phi =k, \,\,\, k\in \R.$$
In order to find some solutions of this equation, we take $k=0$. By this choice, the equation becomes a {\it Bernoulli} ODE.
We make a change of the unknown function, $\psi =\phi^{-2},$
and the Bernoulli ODE becomes a linear ODE $\psi '-\frac{2c}{a}\psi = -\frac{2d}{3a},$
with the solutions
$$\psi(z)=\exp\left({\di\int\frac{2c}{a}dz}\right)\, 
\left[K-\int\frac{2d}{3a}\, \exp\left({-\di\int\frac{2c}{a}dz}\right)dz\right], \,\,\, K\in \R,$$
that is
$$\psi(z)=K\exp\left({\di\frac{2c}{a}z}\right)+\frac{d}{3c}, \,\,\, K\in \R.$$
Since $\phi =\psi^{-\frac{1}{2}}$, we obtain
$$\phi(z)=\frac{1}{\sqrt{K\exp\left({\di\frac{2c}{a}z}\right)+\di\frac{d}{3c}}}, \,\,\, K\in \R.$$
Thus we obtain a family of multitime soliton-solutions of the
multitime Van der Pol PDE ($3^{\prime}$) and we can formulate the next
theorem:

\begin{Th} If we take $\lambda_{\al}$ so as to fix $h, D, C, \lambda$ by the conditions (9),
then we get a family of multitime soliton-solutions 
$$u(x,t)=\frac{1}{\sqrt{K\exp\left({\di\frac{2C^{\ga}\lambda_{\ga}}{h^{\al \be}\lambda_{\al} \lambda_{\be}-1}(x-\lambda _{\al}t^{\al})}\right)+
\di\frac{D^{\ga}\lambda_{\ga}}{3C^{\ga}\lambda_{\ga}}}}, \,\,\, K\in \R.$$
of the multitime Rayleigh wave equation of Van der Pol  type $(3^{\prime})$.
\end{Th}

\section{Stability of multitime Rayleigh solitons}

The multitime Rayleigh PDE is an evolution equation. To show what happen in {\it future multitime} $t$,
we endow the set $\R^m_+$ with the product order. 
Also, suppose $\lambda_\alpha >0$, for each index $\alpha$. The constant vector $\lambda=(\lambda_\alpha)$ controls 
the speed, amplitude, and width of a multitime soliton.

The multitime Rayleigh PDE has important properties of stability, among which there is the following: 
(i) if we specify the initial position $u(x, 0) = u_0(x)$ of a multitime soliton $u(x,t)$ 
at multitime $t = 0$, the equation has a unique solution with that initial data for 
all future multitimes $t > 0$; (ii) if we impose
suitable conditions for the coefficients in a multitime soliton $u(x,t)$, then
$\displaystyle\lim_{||t|| \to \infty}\, u(x,t) = 0$; (iii) let $u(x,t),\,u(x,0)=u_0(x)$ 
be a multitime soliton and $u(x)= ax+b$ be a stationary solution of the Rayleigh PDE; in the space $H^{\frac{1}{2}}$, 
if $||u_0(x)-u(x)||\leq \epsilon$, then $\displaystyle\sup_{||t||}\inf_x\, ||u(x,t)-u(x)|| \leq C\epsilon$, 
under reasonable conditions. 

{\bf Acknowledgements} We wish to thank the referees for their useful comments.
Partially supported by University Politehnica of Bucharest
and by Academy of Romanian Scientists.


\end{document}